\documentclass[12pt]{amsart}
\usepackage[margin=1in,includefoot]{geometry}                % See geometry.pdf to learn the layout options. There are lots.
\usepackage{graphicx}
\usepackage{amssymb}
\usepackage{epstopdf}
\usepackage{natbib}
\usepackage{color}
\begin{document}
%\maketitle

\begin{center}
{\Huge \bf Detecting volcanically produced tori along orbits of exoplanets using UV spectroscopy}\\
%White Paper Subtitle [Optional]
\end{center}

\vskip 3cm

\noindent {\bf Corresponding Author:}\\
Kristina G. Kislyakova -- kristina.kislyakova@univie.ac.at \\
Department of Astrophysics, University of Vienna\\
T\"urkenschanzstrasse 17, 1180 Vienna\\
Austria\\

\noindent Space Research Institute, Austrian Academy of Sciences\\
Schmiedlstrasse 6, 8042, Graz\\
Austria\\

%\noindent {\bf Additional Authors:}
%Luca Fossati (Space Research Institute, Austrian Academy of Sciences), \red{...}

\noindent {\bf Additional Authors:}
Luca Fossati (Space Research Institute, Austrian Academy of Sciences), Denis Shulyak (Max-Planck-Institute f\"ur Sonnensystemforschung), Eike G\"unther (Th\"uringer Landessternwarte Tautenburg, Tautenburg, Germany), Manuel G\"udel (Department of Astrophysics, University of Vienna), Colin P. Johnstone (Department of Astrophysics, University of Vienna), Vladimir Airapetian (Goddard Space Flight Center, NASA, USA), Sudeshna Boro Saikia (Department of Astrophysics, University of Vienna), Allan Sacha Brun (Laboratoire AIM  Paris-Saclay, France), Vera Dobos (Konkoly Observatory, Budapest, Hungary), Kevin France (Laboratory for Atmospheric and Space Physics, University of Colorado, Boulder, USA), Eric Gaidos (Department of Earth Sciences, University of Hawaii, Honolulu, USA), Maxim L. Khodachenko (Space Research Institute, Austrian Academy of Sciences),  Antonino F. Lanza (Istituto Nazionale di AstroFisica, Catania, Italy), Helmut Lammer (Space Research Institute, Austrian Academy of Sciences), Lena Noack (Freie Universit\"at Berlin, Germany), Rodrigo Luger (Center for Computational Astrophysics, Flatiron Institute, New York, USA), Antoine Strugarek (Laboratoire AIM Paris-Saclay, CEA/Irfu Universit$\acute{\rm e}$ Paris-Diderot CNRS/INSU), Aline Vidotto (School of Physics, Trinity College Dublin, Ireland), Allison Youngblood (Goddard Space Flight Center, NASA, USA)

\vfill
\pagebreak
\setcounter{page}{1}
\begin{center}
{\Huge \bf Detecting volcanically produced tori along orbits of exoplanets using UV spectroscopy} \\
%White Paper Subtitle [Optional]
\end{center}
\vskip 1.5 cm

\section{Introduction}

\noindent Significant progress has been made since the discovery of the first exoplanet, 51~Peg~b \citep{Mayor95}, including collecting statistics on occurrence rates, masses, and radii of exoplanets (www.exoplanet.eu). Characterization is the next logical step. The Hubble Space Telescope ({\it HST}) is providing crucial observations for planet atmospheric characterization, particularly at ultraviolet (UV) wavelengths, which are not accessible from the ground. 

%{\it HST} has so far provided the most homogeneous characterization of exoplanetary atmospheres indicating the presence of a great variety of atmospheres with a diversity of features \red{(Sing et al. 2015)}. 
{\it HST} has been extremely effective in characterizing the upper atmospheres of giant exoplanets by observing escaping hydrogen (e.g., \citealp{VM03,Ehrenreich15,Bourrier18}) and minor species present in the upper atmospheres of these planets \citep{Fossati10,Haswell12,VM13}. Characterizing the atmospheres of small rocky planets is more difficult due to their smaller sizes and pressure scale heights compared to those of giant planets. We suggest a new method for characterizing rocky exoplanets, namely, the detection of volcanically-produced tori lying along the planetary orbits. As we show below, {\it HST} is capable of detecting an Io-like plasma torus around a bright M~dwarf in the solar neighborhood. Observations of this kind would be extremely timely as they would combine the unique observational capabilities of {\it HST}, particularly at far-UV (FUV) wavelengths, and new targets discovered by the {\it TESS} satellite. {\it TESS} is a whole-sky survey \citep{Ricker16} that is currently detecting a number of planets orbiting bright (V$<$10 mag) and close enough ($<$50\,pc) systems to enable characterization. Because of the large number of M~dwarfs in the solar neighborhood and the red bandpass and observing windows of {\it TESS}, one can expect that {\it TESS} is likely to detect numerous close-in planets orbiting M dwarfs \citep{Barclay18}. %Additional transiting and non-transiting planets will be found also from the ground by facilities such as {\it TRAPPIST}, {\it NGTS}, {\it CARMENES}, {\it SPIROU}, {\it GIARPS}, and {\it NIRPS}. 
Follow-up {\it HST} observations of these systems, particularly at FUV wavelengths, will give us important insights into atmospheres and even interior compositions of these planets. 

\section{Status of research}

\noindent An exoplanet requires internal heating sources to drive its volcanic activity. Several sources of internal energy are known for planets: radioactive decay, mantle differentiation, (inner) core formation, tidal heating, and induction heating. Due to the compactness of the planetary systems so far detected orbiting late M~dwarfs and to the strong magnetic fields of the host stars, the latter two mechanisms, while insignificant for the Earth, can be very powerful for planets orbiting M dwarfs. Both of them are capable of generating enough heat to produce a global subsurface magma ocean \citep{K17,Dobos19}. 

A sub-surface magma ocean can drive enormous volcanic activity, as is the case for the Jovian satellite Io due to the tidal interaction with other Galilean satellites \citep{Peale79}. Outgassed material is quickly lost to space, and forms a torus around Jupiter, along the moon's orbit, populated mostly by oxygen and sulfur atoms and ions (Fig.~1; e.g., \citealp{Murakami16}). Io's torus has been observed both from space by the UVIS instrument on-board Cassini \citep{Steffl04} and by Voyager~1 \citep{Volwerk18}, and from the ground (e.g., \citealp{Thomas96}). Techniques for the removal of the geocoronal emission produced by neutral hydrogen and oxygen have been only recently developed \citep{Ben-Jaffel13}, which is probably why Io's torus has not yet been observed with {\it HST}.
%--------------------------------------------------
\begin{figure*}
\begin{center}
\includegraphics[width=0.8\columnwidth]{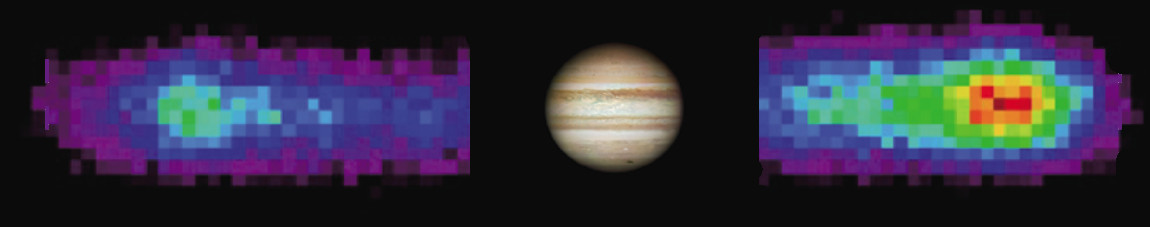}
%\caption{Top: A sketch illustrating formation of a torus due to outgassing of volatiles with their consequent loss to space. Bottom: Observation of the Io plasma torus in SIII line by \textit{Hisaki} satellite \citep{Murakami16}.} 
\caption{Observation of the Io plasma torus in SIII line by \textit{Hisaki} satellite \citep{Murakami16}.}  
\end{center}
\end{figure*}
%--------------------------------------------------

We suggest that similar tori can be produced by exoplanets in close orbit around late M~dwarfs. A dense-enough torus can absorb the stellar light at the position of strong resonance lines of abundant elements, as in the case of the WASP-12 system \citep{Haswell12,Fossati13}. \citet{K18} studied the detectability of absorption signatures by O{\sc i} superposed to the stellar FUV emission triplet at $\lambda\approx$\,1304\,\AA\ taking the HST/STIS E140M observations of the active M~dwarf AD~Leo as a reference for the line strength and width. They have shown that at a spectral resolution of about 50\,000 (e.g., HST/STIS E140M grating) and a signal-to-noise ratio (S/N) of 10, which is currently reachable with {\it HST} for nearby M dwarfs, the torus would be detectable during one \textit{HST} orbit if it had a column density larger than 10$^{13}$\,cm$^{-2}$ (Fig.~2). For comparison, the column density of Io's torus is of the order of 10$^{13}$--10$^{14}$~cm$^{-2}$ \citep{Steffl04}. Planets orbiting M~dwarfs are likely subject to efficient atmospheric escape (e.g., \citealp{Airapetian17,Garcia-Sage17}), which makes formation of an even denser plasma torus likely. Previously, \citet{Demory16} have found indications of volcanism on 55\,Cnc\,e by observing secondary eclipses and phase curves at mid-infrared wavelengths. \cite{Ridden-Harper16} have detected a tentative sodium signal for 55\,Cnc\,e, which could potentially also have origin in volcanic activity. These studies suggest that volcanism can in principle be detectable on exoplanets.

%--------------------------------------------------
\begin{figure*}
\begin{center}
\includegraphics[width=0.6\columnwidth]{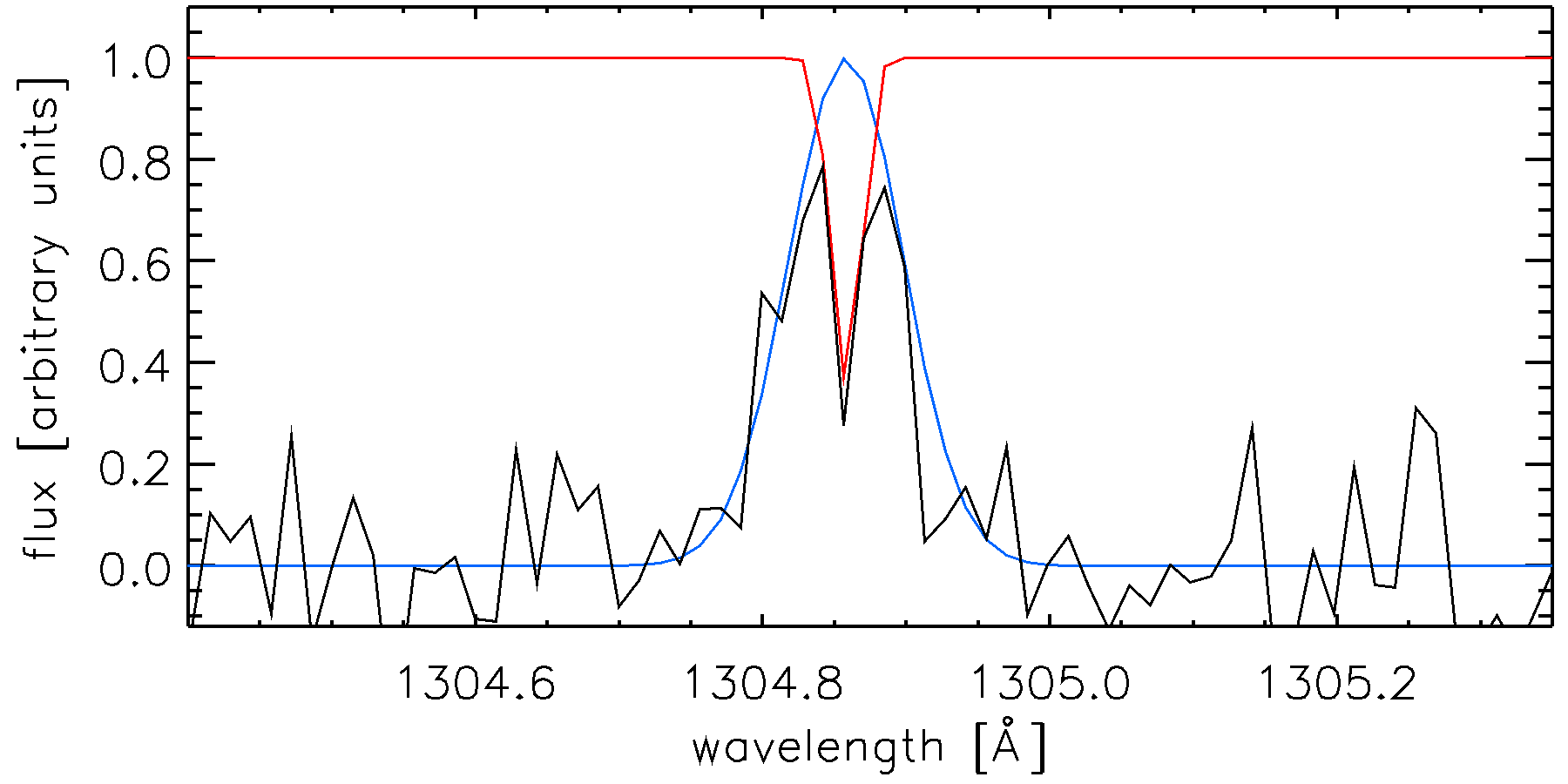}
\caption{Synthetic Gaussian profile of the O{\sc i} line at 1304.858\,\AA\ (blue) superposed to a synthetic Voigt absorption feature (red)
computed assuming a column density of 10$^{13}$\,cm$^{-2}$. The final emission line (black), which includes the torus absorption, was computed assuming a signal-to-noise ratio of 10. The spectral resolution is that of the E140M HST/STIS grating. From \citet{K18}.}        
\end{center}
\end{figure*}
%--------------------------------------------------

{\it HST} follow-up observations of planets orbiting M~dwarfs can be used for detection of volcanic activity on these planets. Both tidal and induction heating are the most effective in planets orbiting late-type M~dwarfs, due to compactness of these systems and strong stellar magnetic fields (e.g. \citealp{Shulyak19,Kochukov19}). However, these stars are also very dim, which requires very long integration times for observations. We suggest to focus on rocky planets orbiting M~dwarfs of spectral types M4--M5, as they are still sufficiently bright, but also still likely to have planets with a sub-surface magma ocean due to tidal or induction heating. A possible target already available is the TRAPPIST-1 system. TTVs that are detected are indicative of tidal effects, including heating \citep{Luger17}. This system has been observed by the \textit{HST} \citep{deWit18}, however, one needs a brighter star to allow for the characterization of a torus. Therefore, a bigger and hotter star of a spectral class M4-M5 with a close-in rocky planet would be an ideal target. Observations of such tori have never been attempted before as only recently {\it TESS} has given us the possibility to find adequate targets for these novel observations.

\section{Science goals of a possible UV observational study}

\noindent \textbf{Planet atmospheric composition and escape - } detecting a torus along the orbit will help us constraining the intensity of escape processes for planets orbiting M~dwarfs. The torus will also enable us to identify the main constituents of the atmosphere.

\smallskip 

\noindent \textbf{Detection of exoplanet volcanic activity - } detecting a plasma torus would confirm the presence of active volcanoes on exoplanets. The material forming the torus should be constantly replenished by outgassing from the planetary interior, which in turn is supplied by the planet's volcanic activity. 

\smallskip 

\noindent \textbf{Characterization of planetary interiors - } {\it HST} observations of the torus will uniquely enable us to constrain the internal composition of exoplanets. This is because the composition of the outgassed material is directly connected to the composition and redox state of the planetary mantle \citep{Gaillard14}. Therefore, observations of this kind would give us the unique opportunity to reveal key properties of planetary interiors, which would be difficult or even impossible to obtain in any other way.

\section{Conclusions}

\noindent The {\it TESS} mission is going to play a pivotal instrumental role in our future understanding of planets, particularly by providing numerous low-mass, transiting planets in close orbit to bright and nearby stars. Ultraviolet observations conducted with {\it HST} have successfully demonstrated their capability of detecting atomic and ionic species escaping from the atmospheres of giant planets. Furthermore, realistic predictions indicate that the same is going to be possible also for the tori of rocky planets in close orbit to nearby M~dwarfs. {\it HST}'s remaining operational time is limited and it is of crucial importance to make best use of its capabilities, particularly at UV wavelengths, to gain as much observational material as possible to inform the science and guide the requirements of the next generation of space telescopes with UV capability.

\vfill
\pagebreak
%{\bf \noindent References:}\\

\begin{small}
\bibliographystyle{apj}
\bibliography{./references_WhitePaper}
\end{small}

\vfill
\end{document}